\documentclass[useAMS,usenatbib,usegraphicx]{mn2e}
\usepackage{aas_macros,times}

\title[ROLES I]
  {A spectroscopic measure of the star-formation rate density in dwarf galaxies at z
  $\sim$ 1} 
\author[G. Davies et al.]
  {G.T.~Davies$^1$,  David G.~Gilbank$^2$\thanks{Email: dgilbank@astro.uwaterloo.ca}, Karl~Glazebrook$^3$, Richard G.~Bower$^1$, 
  \newauthor
  I.K. ~Baldry$^4$,   Michael L.~Balogh$^2$,
  G.K.T.~Hau$^{1,3}$, I.H.~Li$^3$, P. ~McCarthy$^5$, S.~Savaglio$^6$
\newauthor\\
  $^1$Institute for Computational Cosmology, Department of Physics, University of Durham, South Road, Durham, DH1 3LE, UK\\
  $^2$Department of Physics and Astronomy, University of Waterloo, Waterloo, Ontario, Canada N2L 3G1\\
  $^3$Centre for Astrophysics and Supercomputing, Swinburne University of Technology, P.O. Box 218, Hawthorn, VIC 3122, Australia\\
  $^4$Astrophysics Research Institute, Liverpool John Moores University, Twelve Quays House, Egerton Wharf, Birkenhead CH41 1LD, UK\\
  $^5$Carnegie Observatories, 813 Santa Barbara Street, Pasadena, California, 91101 USA\\
  $^6$Max-Planck Institute for Extraterrestrial Physics, Giessenbachstrasse, 85748 Garching bei Munchen, Germany\\
}
\date{\today}

\pagerange{\pageref{firstpage}--\pageref{lastpage}} \pubyear{2006}

\def\LaTeX{L\kern-.36em\raise.3ex\hbox{a}\kern-.15em
    T\kern-.1667em\lower.7ex\hbox{E}\kern-.125emX}

\def\oii{[{\sc OII}]}

\def\k{$K_{AB}$}

\def\ms{$\log(M_*/M_\odot)$}

\def\Msun{M_{\odot} }

\def\gsim{\mathrel{\raise0.35ex\hbox{$\scriptstyle >$}\kern-0.6em
\lower0.40ex\hbox{{$\scriptstyle \sim$}}}}
\def\lsim{\mathrel{\raise0.35ex\hbox{$\scriptstyle <$}\kern-0.6em
\lower0.40ex\hbox{{$\scriptstyle \sim$}}}}

\begin{document}

\label{firstpage}

\maketitle

\begin{abstract}
We use a $K$-selected (22.5 $<$ \k $<$ 24.0) sample of dwarf galaxies (8.4 $<$ log($M_*/M_{\odot}$) $<$ 10) at 0.89$<$z$<$1.15 in the {\it Chandra} Deep Field South (CDFS) to
measure their contribution to the global star-formation rate density
(SFRD), as inferred from their \oii~flux.  
By comparing with \oii-based studies of higher stellar mass galaxies, we robustly measure a turnover in the \oii~luminosity density at a stellar mass of $M\sim10^{10}M_\odot$.  By comparison with the \oii-based SFRD measured from the {\it Sloan Digital Sky Survey} we confirm that, while the SFRD of the lowest-mass galaxies changes very little with time, the SFRD of more massive galaxies evolves strongly, such that they dominate the SFRD at $z=1$.
\noindent
\end{abstract}


\section{Introduction}
\label{sec:introduction}
Remarkable progress has been made over the last ten years in putting together an increasingly detailed picture of galaxy evolution since
$z<4$.   In particular, having established with some accuracy the star-formation
history of the Universe \citep[e.g.][]{Lilly,HB06,Reddy}, the next objective is to
establish how these stars were assembled over time
\citep[e.g.][]{dick03,cons07,bell07,cowi08,March08}.  This mass-assembly 
history is, in principle, an observable quantity that can provide a
robust, direct constraint on theoretical models (e.g., \citealt{bowe06}).

One of the most generic predictions of all galaxy formation models is that the total
mass in the Universe, dominated by cold dark matter (CDM, \citealt{blum84}), assembles
by building up progressively larger structures with time
\citep[e.g.][]{wf91}.  Observations
have long shown that the most massive galaxies 
today actually have the {\it oldest} stellar populations
\citep[e.g.][]{gall84,ble,vD98,Nelan,Smith,rettura}, but this alone does
not pose much difficulty for theory if these massive galaxies were
assembled early from smaller lumps of matter in which stellar populations
were already established.  More puzzling have been direct
observations of high-redshift galaxies, which show that the majority of
massive galaxies were already in place by $z=1$, and that they stopped forming
new stars sooner than galaxies of lower mass (\citealt{cowi96},
\citealt{june05} (hereafter J05), \citealt{font04}, \citealt{bundy},
\citealt{moba08}, \citealt{taylor}).

Thus, it is of key interest to obtain a direct measurement of star-formation rate as a
function of stellar mass in galaxies at different redshifts. To date, high-redshift measurements have been limited to the most
massive or the most highly-star forming galaxies. Recent near-infrared
selected spectroscopic surveys such as the Gemini Deep Deep Survey
(GDDS, \citealt{abra04}) and K20 \citep{font04} have pushed as deep as
$K_{AB}\simeq 22.5$. These are desirable as the $K$-band allows a
clean selection to be made on approximate stellar mass out to
high-redshifts, and stellar mass is a relatively robust quantity to compare with
simulations \citep[e.g.][]{March08}. GDDS and K20 select $>10^{11}\Msun$ (stellar mass)
galaxies to $z\simeq 2$ and $>10^{10}\Msun$ galaxies at $z\simeq
1$. Spectroscopic surveys serve to provide accurate redshifts and also
to measure fluxes in nebular lines such as [OII] and H$\alpha$ which
can be used to estimate star-formation rates. However, spectroscopy is
generally not attempted for fainter continuum objects due to the much
longer integration times required to assemble a large sample of
objects. This results in a major limitation to earlier work such as
J05 --- the low mass bins at high-redshifts are grossly incomplete,
even though this is a relatively deep survey. Other spectroscopic works
such as the Deep Extragalactic Evolutionary Probe (DEEP2,
\citealt{davi03}, \citealt{will06}) and VIRMOS-VLT deep survey (VVDS,
\citealt{lefe03}) are even shallower and only probe the most massive
systems. The only studies of star-formation in high-redshift, low-mass
systems either rely on photometric redshifts without spectral
information (e.g. the Great Observatories Origins Deep Survey, GOODS,
\citealt{dick03} and the MUltiwavelength Survey by Yale-Chile, MUSYC \citealt{taylor}) or use random Gamma Ray Burst events to select the
spectroscopic targets \citep{sava08}.  Thus, it is still an open
question whether low mass galaxies have always had high SFRs compared
with their higher mass counterparts (as
observed locally), or whether the bulk of star-formation has actually
progressed from high mass systems to low mass systems with increasing
cosmic time.

In this Letter we report the first results from the `Redshift One
LDSS-3 Emission line Survey' (ROLES) which utilises a novel approach to obtain a census of the star-formation rates in z$\sim$1 galaxies an order of magnitude lower in stellar mass than previously studied with spectroscopic techniques.   The availability of \oii~observations locally down to equivalently low stellar masses, and z$\sim$1 \oii~measurements for higher stellar masses, mean that we can consistently compare \oii~as a function of redshift and stellar mass.  We adopt a flat cosmology with ($H_0$, $\Omega_m$) =
(70, 0.3).  All our magnitudes are on the AB system unless otherwise
noted, with $K_{\rm AB}=K_{\rm Vega}+1.87$.

\section{Survey Design, Observations and Data}
\label{sec:method}

ROLES utilises fields with deep $K$-band imaging and photometric redshifts in order to pre-select likely low stellar mass ($K$-faint, $22.5 < K \le 24.0$) galaxies at z$\sim$1.  We target these objects with multi-object spectroscopy (MOS) using LDSS-3 on the 6.5-m Magellan telescope.  We use the \oii$\lambda$3727 emission line to both obtain a spectroscopic redshift and to estimate the star-formation rate (SFR) of each object.  In order to increase our observing efficiency further, we use a custom band-limiting filter spanning 7040\AA~to 8010\AA~FWHM.  This restricts our wavelength range for observing \oii~to 0.889$<z\le1.149$ and we preferentially target galaxies with photometric redshifts compatible with this range.   With this approach we will obviously not obtain redshifts for galaxies without emission lines, but these do not contribute to the star-formation rate density of the Universe.  In this way, we can efficiently build a sample which is stellar mass selected and complete to a given (unobscured) SFR limit.

We use LDSS-3 in nod-and shuffle \citep{ns} mode (N\&S) to obtain the best possible sky subtraction, and typically place $\sim$200 0.8\arcsec wide slits over the $\sim$8.2 arcmin field of view.  Total exposure times for each mask are typically four hours, and here we present results from one of our fields (the first 491 slits, resulting in 171 redshifts, 64 of which are at z$\sim$1) from an ongoing project to observe sources in two deep fields.  The flux limit, conservatively set to the depth of our shallowest mask, can be seen in Fig.~1.  Details of our full sample, observing setup and data reduction will be presented in a forthcoming paper.  The data were reduced to 2D spectra using a combination of custom-written routines and 
the {\sc COSMOS2}\footnote{see
http://users.ociw.edu/oemler/COSMOS2/COSMOS2.html.} software in a manner standard for N\&S spectroscopy (see also \citealt{gregthesis}).

\section{Method}
We identify emission line features in the 2D spectra in the following manner.  We propagate a detailed estimate of the noise of each pixel through our reduction process.  We convolve the final reduced image with a kernel of the typical profile of our emission lines.  We compare local enhancements in the smoothed signal frame with the expected noise (also making allowance for the contribution due to continuum emission) and retain peaks of more then 4.5$\sigma$ significance (this limit minimises obvious spurious detections, whilst maintaining high completeness, as confirmed by independent visual inspection).  Most of our \oii~detections are necessarily single line detections (in a few cases \oii~is supported by [NeIII]$\lambda$3869), so we use redshift probability distribution functions (from the FIREWORKS dataset, \citealt{wuyt08}, kindly provided by S. Wuyts) to identify lines most consistent with being \oii.  Comparison with multiple-line redshifts in our own data (mostly at z$\sim$0.5), and with a few objects in common with public spectroscopy \citep{vanz08}, shows that this works well.  In most cases, the probability either overwhelmingly favours \oii~or a different line, and in practice there is little ambiguity.

\begin{figure}
	{\centering
	\includegraphics[width=85mm,angle=0]{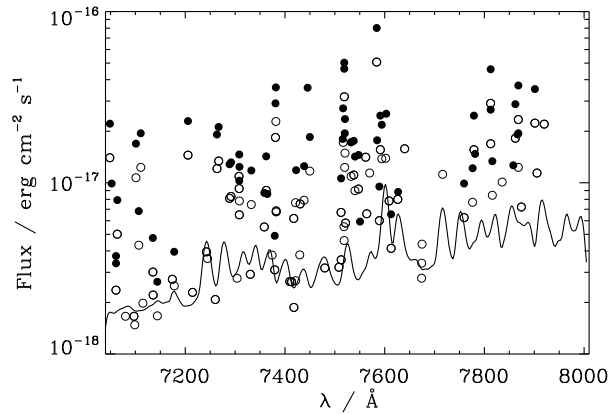}
	\caption{Flux versus wavelength for all detected lines (open circles) and lines most-likely \oii~detections (filled circles).  The solid line is the 4.5$\sigma$ flux limit derived
	from our average noise estimate in the shallowest detection image.  }
	\label{fig:flux_lambda}}
\end{figure}

\begin{figure*}
	{\centering
	\includegraphics[width=135mm,angle=0]{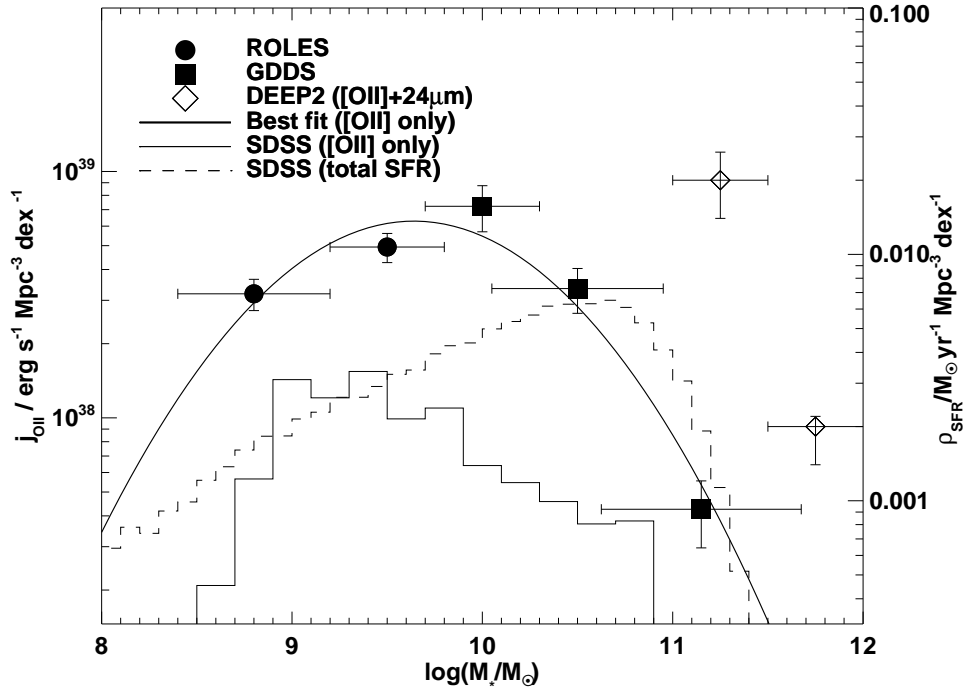}
	\caption{\oii~luminosity density, $j_{\rm OII}$, per log(stellar mass), and the equivalent SFRD under the simple assumptions stated in the text,	versus stellar mass at $z\sim1$.  Filled circles show measurements from \oii~only (ROLES and GDDS) and open circles show the SFRD from \oii+24$\mu$m data.  The solid curve shows a quadratic fit for illustrative purposes.  Solid and dashed histograms show the local \oii-derived and total SFRD estimates respectively, using SDSS data.}
	\label{fig:sfrd}}
\end{figure*}

We measure fluxes for lines identified as \oii~using a simple rectangular aperture around the peak of the emission in the original (i.e., unsmoothed) 2D spectrum.  Our objects are relatively small compared with the size of the slit and we estimate from ACS images convolved to our ground-based seeing that we miss no more than 20\% of the light from each galaxy.  For flux calibration, we used spectrophotometric standard stars to correct the shape of the instrument response.  We set the normalisation by comparison with other flux-calibrated surveys having objects in common with our sample.  We have 13 galaxies with \oii~fluxes in common with the public ESO spectroscopy of \citet{vanz08}, and find a consistent offset (for all three masks) with $\sim$30\% scatter when comparing our fluxes with their data (which should be close to total fluxes for our compact objects).  The public fluxes should be reliable for these objects, as determined from cross-checks against broad-band photometry for a large sample of continuum objects (E. Vanzella, priv. comm.).  In addition, we have three galaxies in common with the IMAGES survey \citep[][flux-calibrated spectra kindly provided by H. Flores and C. Ravikumar]{Ravikumar:2007nn} and these show good agreement with the normalisation calculated from the public ESO spectroscopy.

We show our line detections in Fig.~\ref{fig:flux_lambda}.  The plot shows the measured flux and wavelength of each significant line detected.  Filled circles indicate those where the line is identified as \oii~on the basis of photometric redshift (or multiple emission lines).  The solid line shows a representative flux limit as determined from the typical noise spectrum of the shallower of our three masks.

We compute the \oii-luminosity density (LD) using the $1/V_{max}$ method after correcting for spectroscopic completeness.  $V_{max}$ was calculated by estimating the volume over which both the ($k$-corrected) $K$-band magnitude and \oii~flux would be above our detection limit at each infinitesimal redshift increment.\footnote{In calculating the \oii~flux limit we used the detailed noise spectrum shown in Fig.~\ref{fig:flux_lambda}. Adopting a constant nominal flux limit instead (i.e. ignoring the volume lost behind individual lines) makes the volume 20\% larger.}  For reference, a galaxy which is visible at all
redshifts (from $0.889 < z\le 1.149$) in ROLES would be drawn
from a volume of $3.2 \times 10^4$ Mpc$^3$.  

We convert \oii~luminosity to SFR using a simple conversion (as is commonly used in the literature, e.g., J05), and explore the uncertainty in this conversion in \S4.  This uses the standard SFR(H$\alpha$) conversion of \cite{kenn98} assuming (\oii/H$\alpha)_{obs}=0.5$, $A_{H\alpha}=1$ and a \citet{bald03} (hereafter BG03) IMF.  Throughout this Letter we convert all SFRs presented to the BG03 IMF.  Since we are primarily interested in extending the GDDS results to lower stellar masses, we adopt exactly their prescription for measuring SFR and discuss systematic effects with mass in the next section. The faintest \oii~flux we could record (4.5$\sigma$) at the redshift limits of our survey corresponds to SFRs of 0.34--0.74 M$_\odot$yr$^{-1}$ (assuming one magnitude of extinction).

We calculate completeness corrections by constructing redshift distributions from summing the photometric redshift probability distributions (or spectroscopic redshifts, where available) for each galaxy in the photometric catalogue within the area targeted for spectroscopy.  We construct the same sum for the galaxies actually targeted for spectroscopy, and the ratio of the latter to the former within our redshift limits gives the spectroscopic completeness.  For the data presented here, the overall completeness is $\sim$60\%, with a weak $K$ magnitude dependence falling to $\sim$40\% in the faintest half-magnitude bin.

Stellar masses are estimated from the $UBVRIJHK_s$ catalogue of \citet{moba04} (kindly provided by B. Mobasher and T. Dahlen), using the technique described in \citet{glaz04} to fit the SED at the spectroscopic redshift we measure.   The uncertainties associated with the stellar mass fitting are $\sim$0.2 dex \citep{glaz04}.

\section{Results and Discussion}

Fig.~\ref{fig:sfrd} shows our estimate of the \oii-LD, per logarithmic stellar mass bin in two mass bins.  
We split our sample of 64 ROLES galaxies into these two bins, divided at \ms $=9.2$, to give  comparable numbers of galaxies in each bin (24 and 40 in the low and high mass bins respectively).  Our error bars include a contribution from a reasonable calibration uncertainty (30\%, the approximate scatter from our comparison with the flux measurements from other surveys).\footnote{We have also estimated a reasonable upper limit based on lowering the significance threshold for which we accept emission lines to a level which clearly introduces many false positives.  The shift due to this moves the data up by approximately the plotted 1$\sigma$ errors.  We estimate the effect of cosmic variance at $\sim$40\% (not included in the plotted error bars) using \citet{some04} for this density of objects.  The full ROLES dataset will allow us to estimate this from a comparison of different fields.}  To this plot, we add the results of the \oii-LD from GDDS data\footnote{We rebin the data (S. Juneau, priv. comm.) slightly in mass from the values used in J05, so that the bins are less incomplete.}  (J05) for galaxies of higher stellar mass.  The combination of these two datasets presents a very clear picture of the mass-dependence of \oii-LD at this redshift.  At a lookback time of around 8 Gyr, the \oii-LD of the Universe was dominated by high stellar mass galaxies, and a turnover in the \oii-LD occurs at \ms$\sim$10.0.  A simple quadratic fit, for illustrative purposes, would show a peak around \ms$\sim$9.5.  
Converting the LD to SFRD under our simple model, we measure an integrated SFRD in z$\sim$1 dwarf galaxies (8.4 $<$ \ms $\le$9.8) of $\rho_{SFR} = (4.8\pm1.7) \times 10^{-3}$ M$_\odot$ yr$^{-1}$ Mpc$^{-3}$.  For the first time, we have detected the turnover in the \oii-LD/SFRD, showing that the contribution of lower mass galaxies, $\lsim10^{9}$M$_\odot$, declines.

We construct a local comparison sample from SDSS data by matching the NYU-VAGC \citep{Blanton:2005pk} sample to  \oii~flux measurements and stellar masses from the Garching DR4 release\footnote{see: http://www.mpa-garching.mpg.de/SDSS/DR4/} \citep{brin04}, again using the $1/V_{max}$ method. In order to cleanly sample the \oii$\lambda3727$ line, we restrict the redshift range of the \oii~sample to $0.032<z\le0.050$.  We use the same conversion from \oii-luminosity to SFR as at high redshift, and apply an aperture correction to correct for flux lost outside the fibre using the ratio of the $g$-band fibre magnitude to $g$-band Petrosian magnitude from the imaging data.  The local \oii-determined SFRD is shown as the solid histogram in Fig.~\ref{fig:sfrd}.  
It is clear that the \oii~SFRD of low mass galaxies at z$\sim$1 is comparable to that observed today and that most of the difference is due to a shift in the turnover from high to low mass galaxies with increasing cosmic time.  

It is well known that the dependence of \oii~luminosity on the underlying SFR is sensitive to the effects of dust and metallicity (e.g., \citealt{jans01}).  The conversion we adopt has been determined empirically from local values \citep{kenn98} for massive galaxies.  It might be expected that the relationship between \oii~luminosity and SFR evolves over the redshift range from z$\sim$0--1.  \citet{tres02} showed for a modest size sample (30 galaxies) that the ratio of H$\alpha$/\oii~remained constant to within a factor of 2, but with significant uncertainty on the scatter, over this redshift range.  For simplicity, in this Letter we have adopted a single conversion for all our SFR measurements.  It is thus straightforward to convert all our measurements back to \oii-LD and/or adopt a different SFR calibration, if desired. Given the form of the mass--metallicity relation (e.g., \citealt{sava05}, \citealt{cowi08}) and the metallicity dependence of the \oii~luminosity (e.g., \citealt{kewl04}), the systematic error in using a single metallicity to estimate \oii-inferred SFR is that we will {\it underestimate} the SFR for high stellar mass (high metallicity) galaxies and {\it overestimate} the SFR for low mass (low metallicity) objects.  This only strengthens our result that the SFRD declines towards lower mass galaxies at z$\sim$1.  

Our assumption of constant dust-reddening likely leads to a similar systematic error.  \citet{kewl04} showed from a local galaxy sample that such an assumption leads one to systematically underestimate the SFR at high SFRs and overestimate the SFR at low SFRs.  This trend also seems to hold for z$\gsim$1 galaxies  \citep{adel00}.  Again, the direction of this trend only strengthens our result of a turnover in the SFRD towards lower mass galaxies.  The effects of dust obscuration can be estimated by, for example, looking at the luminosity in the mid-IR, where the emission traces the re-radiated  light, due to star-formation, absorbed by dust.  \citet{cons07} used 24$\mu$m {\it Spitzer} observations combined with \oii~measurements from DEEP2 optical spectroscopy, to attempt to correct \oii-inferred SFRs for the effects of dust obscuration.  Their \oii$+$24$\mu$m SFRs are shown as the open diamonds in Fig.~\ref{fig:sfrd}, and do indeed suggest that the SFR estimates for high-mass galaxies
might be underestimated by \oii-only measurements.  Correcting observed SFRs to something approximating {\it total} SFRs in this way, it can be seen that such corrections can be significant, at least for the high mass galaxies (\ms$\gsim$11) sampled by the \citet{cons07} data.  Deep 24$\mu$m data also exist for the CDFS field presented here.  Unfortunately, since the surface density of our low mass galaxies is much higher than that of higher mass galaxies, the majority of our sources are limited by confusion and extracting meaningful 24$\mu$m luminosities is likely not possible.  SFR estimates using other indicators such as 2000\AA~flux (and 24$\mu$m emission, where possible) will be presented for ROLES galaxies in a future paper. We note that a potentially small contamination from AGN may exist in our data (although none of our galaxies is X-ray detected). This would also only lower the \oii-LD in ROLES.

Finally, we overplot the total SFRD (primarily using H$\alpha$, but including mass-dependent extinction and metallicity effects) measured locally from SDSS data \citep{brin04} in Fig.~\ref{fig:sfrd}.  The \citet{brin04} SFR peak occurs at a much higher mass than that of the simple \oii-inferred estimate (\ms$\sim10.5$ versus \ms$\sim9.0$).  Given that the \oii~SFR traces the H$\alpha$ SFR, at least on average \citep{jans01,tres02}, the likely  cause of this discrepancy is the use of a single dust extinction factor for the \oii~estimate.  The individual fits of extinction to each galaxy used by \citet{brin04}  can vary by several magnitudes. Over the relatively small mass range used in ROLES, the dust extinction probably does not vary systematically by a large amount, but in looking at the SFRD over a wider mass range (e.g., ROLES$+$GDDS), the comparison at low-redshift shows that some care must be taken in interpreting accurately the position of the peak of the SFRD.  Indeed, as we will show in future work,  a relatively simple model for mass-dependent extinction goes most of the way to reconciling the z$\sim$0 \oii-inferred and total SFRDs; however, it is unclear if this is applicable at z$\sim$1.  Given that the conversion from \oii~to total SFRDs locally can shift the peak by two orders of magnitude in stellar mass, we are cautious about making quantitative statements regarding the evolution of the peak of the SFRD, but our result that the SFRD at z$\sim$1 declines towards the lowest mass galaxies is robust to any such uncertainties.

\section{Conclusions}
\label{sec:conclusions}

We have presented the first spectroscopic measurement of the \oii-LD in dwarf galaxies at z$\sim$1, using the \oii $\lambda$3727 line. Comparing the \oii-LD with higher mass galaxies, we find that the contribution to the \oii-LD at this redshift is declining below a mass of \ms$\sim$10.  Under the simple assumption of constant dust extinction (as used in J05 and other works), we convert L(\oii) to SFR and measure a SFRD of $(4.8\pm1.7) \times 10^{-3}$ M$_\odot$ yr$^{-1}$  Mpc$^{-3}$ from 64 galaxies with stellar masses of 8.4 $<$ \ms $\le$9.8.  Comparing the z$\sim$1 \oii-LD with the local value suggests that the overall SFRD in low mass (\ms$\lsim$9.5) galaxies has remained roughly constant over this period, whilst the bulk of the SFRD has shifted from high mass to low mass galaxies with increasing cosmic time.  This is one manifestation of the picture generally referred to as ``downsizing"  \citep{cowi96} of star-formation, although different workers have different definitions of this (e.g., downsizing could equally well refer to a shift in the overall normalisation of the SFRD--mass plot and/or a shift in the position of the peak - we find the latter here).

These are the first results from the Redshift One LDSS-3 Emission line Survey (ROLES).  The entire ROLES dataset for this redshift range comprises approximately six times as many objects as presented here and spans two fields to assess the impact of cosmic variance.  NIR spectroscopic follow-up to obtain H$\alpha$ SFRs for many of the galaxies is underway.

\section*{Acknowledgements}
\label{sec:acknowledgements}
This paper includes data gathered with the 6.5 metre Magellan
Telescopes located at Las Campanas Observatory, Chile.  We thank LCO and the OCIW for the allocation of time to this project as part of the LDSS3 instrument project.  We thank T. Dahlen,  H. Flores, S. Juneau, B. Mobasher, C. Ravikumar, E. Vanzella and S. Wuyts  for graciously providing data and for useful discussions. We thank the referee for a very thorough report which led to a clearer presentation of our results.  GTD acknowledges the receipt of an
STFC PhD studentship.  Karl Glazebrook and I-H Li acknowledge
financial support from Australian Research Council (ARC) Discovery Project DP0774469. Karl
Glazebrook and Ivan Baldry acknowledge support from the David and
Lucille Packard Foundation. MLB acknowledges support from the province of Ontario in the form of an Early Researcher Award.  GKTH thanks ARC for financial support.

\label{lastpage}


\end{document}